\documentclass[aps,10pt,singlespacing,nofootinbib,preprintnumbers,prd,twocolumn,showpacs]{revtex4-1}
\usepackage{amsmath}
\usepackage{bm}
\usepackage{times}
\usepackage[pdftex]{graphicx}

\usepackage{color,graphicx}
\usepackage{amssymb} 
\usepackage{bm}
\usepackage{braket}
\usepackage{slashed}
\usepackage{hyperref}
\usepackage{latexsym}
\usepackage{bm}

\usepackage[english]{babel}

\definecolor{nicered}{rgb}{0.7,0.1,0.1}
\definecolor{nicegreen}{rgb}{0.1,0.5,0.1}

\def \eff{{\text{eff}}}
\newcommand{\qq}{(q^2)}
\newcommand{\nn}{\nonumber}
\newcommand {\E}[1]{\times 10^{#1}} % exponent notation
\newcommand {\e}[1]{\mathrm{~#1}} % units
\newcommand{\mc}[1]{\mathcal{#1}}

\newcommand{\mrm}[1]{\mathrm{#1}}

\renewcommand{\Re}[0]{\mrm{Re}}

\definecolor{Red}{rgb}{1.,0.,0.}

\definecolor{Blue}{rgb}{0.,0.,1.}

\usepackage{array}
\usepackage{multirow}
\usepackage{color,colortbl}

\hypersetup{colorlinks,citecolor= nicegreen,linkcolor= nicered}
\hypersetup{colorlinks=false}
\begin{document}
\title{Model independent constraints on leptoquarks from $b\to s\ell^+ \ell^-$ processes}

\author{Nejc Ko\v snik}
\email[Electronic address:]{nejc.kosnik@ijs.si}
\affiliation{Laboratoire de l'Acc\'el\'erateur Lin\'eaire,
Centre d'Orsay, Universit\'e de Paris-Sud XI,
B.P. 34, B\^atiment 200,
91898 Orsay cedex, France}
\affiliation{J. Stefan Institute, Jamova 39, P. O. Box 3000, 1001 Ljubljana, Slovenia}

\begin{abstract}
  We list all scalar and vector leptoquark states that contribute to
  the $b \to s \ell^+ \ell^-$ effective Hamiltonian. There are
  altogether three scalar and four vector leptoquarks that are
  relevant. For contribution of each state we infer the correlations
  between effective operators and find that only two baryon
  number-violating vector leptoquarks give rise to scalar and
  pseudoscalar four-fermion operators, whereas the scalar states can
  contribute to those operators only when two states with same charge
  are present. We bound the resulting Wilson coefficients by imposing
  experimental constraints coming from branching fractions of $B \to K
  \ell^+ \ell^-$, $B_s \to \mu^+ \mu^-$, and $B \to X_s \mu^+ \mu^-$
  decays.
\end{abstract}

\preprint{LAL-12-183}
\maketitle

\section{Introduction}
The $b \to s \ell^+ \ell^-$ induced processes have been recognized as
very important probes of the Standard Model and new physics. Rare
decay $B_s \to \mu^+ \mu^-$ has been subject to intensive experimental
efforts~\cite{Abazov:2010fs,*Aaltonen:2011fi,*Chatrchyan:2011kr,*PhysRevLett.108.231801}
at Fermilab and LHC and currently the upper bound on the branching
ratio has been set slightly above the Standard Model~(SM)
prediction. Increasing statistics in this decay mode at the LHC will
soon allow to probe the SM prediction
directly~\cite{Buras:2003td}. Exclusive $B \to K^{(*)} \ell^+ \ell^-$
and inclusive $B \to X_s \ell^+ \ell^-$ decays with $\ell = e, \mu$
offer many different observables to be confronted against the
theoretical predictions. Their studies at the $B$-meson
factories~\cite{belle-incl,babar-incl,:2012vw} and at the LHCb
experiment~\cite{Aaij:2012cq} indicate that all observables are,
within relatively large error bars, compatible with the
predictions of the SM~\cite{Alok:2010zd,*Drobnak:2011aa,*Bobeth:2011nj,*Beaujean:2012uj,*Mahmoudi:2012un,*Altmannshofer:2012ir}.

The leptonic branching fraction, $\mrm{Br}(B_s \to \mu^+ \mu^-)$, is
very sensitive to physics beyond the SM where scalar or pseudoscalar
four-fermion operators are present, namely, such contributions are
helicity-enhanced with respect to the SM amplitude. Complementary
information on those operators can be extracted from the spectrum of
semileptonic $B \to K \ell^+ \ell^-$ decay. Indeed, the leptonic and
semileptonic decay widths depend on orthogonal combinations of
(axial-)vector current and (pseudo)scalar four-fermion
operators~\cite{Becirevic:2012fy}. Size of the vector and axial-vector
current operators can also be assessed by studying the transverse
asymmetries in $B \to K^* \ell^+ \ell^-$
decay~\cite{Kruger:2005ep,*Becirevic:2011bp}.

Scalar and pseudoscalar operators are present in new physics~(NP)
models where a color- and charge-neutral scalar particle produces the
lepton pair, as is the case in supersymmetric extensions of the
SM. Another possibility to generate $b\to s \ell^+ \ell^-$ at short
distances is an exchange of a color triplet particles that couple to a
lepton-quark pair. Such leptoquark states have spin either 0 and 1 and
are present in Grand Unified
Theories~\cite{Georgi:1974sy,*PhysRevLett.64.619,*Dorsner:2005fq},
Pati-Salam models~\cite{Pati:1974yy}, composite
scenarios~\cite{Schrempp:1984nj,*Gripaios:2009dq}, or technicolor
models~\cite{Kaplan:1991dc}. However, since a leptoquark naturally
generates Fierzed operators of the form $(\bar s \Gamma \ell)(\bar
\ell \Gamma b)$, the scalar operators,
\begin{equation}
   (\bar{s} P_{L(R)} b)(  \bar{\ell} \ell)\,,\qquad  (\bar{s} P_{L(R)} b)(  \bar{\ell} \gamma_5\ell)\,,
\end{equation}
cannot be identified with exchanges of a scalar leptoquarks. In a
similar way, a vector leptoquark exchange does not necessarily induce
vector current operators.

\begin{figure}[!hbtp]
  \centering
  \begin{tabular}{cccc}
    \includegraphics[width=0.22\textwidth]{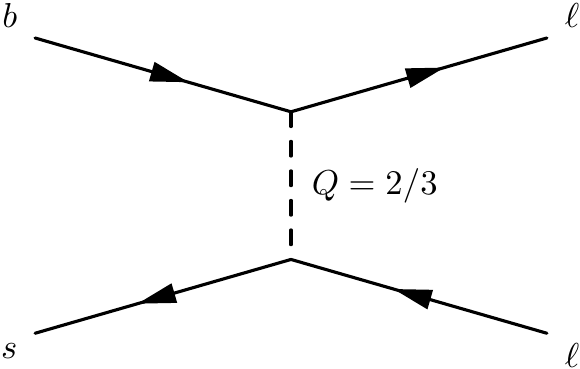}
    && &    \includegraphics[width=0.22\textwidth]{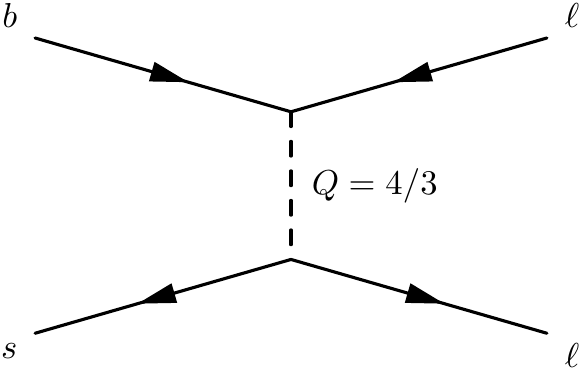}
  \end{tabular}
  \caption{Two possible charges of a leptoquark in $b \to s \ell^+
    \ell^-$ diagram.}
  \label{fig:flow}
\end{figure}

Leptoquarks have been studied extensively in the literature. For early
model independent studies see
e.g.~\cite{Buchmuller:1986zs,*Leurer:1993em,*Davidson:1993qk,*PhysRevD.56.5709},
while for some recent works see~\cite{Alikhanov:2012kk,*Saha:2010vw,*Dighe:2010nj,*Carpentier:2010ue,*Bobeth:2011st,*Dorsner:2011ai}. In
this work we complement the SM with a single leptoquark state and
assume all other degrees of freedom lie substantially higher above the
electroweak scale. The tree-level contributions to $b\to s \ell^+
\ell^-$ due to a single colored particle exchange present a very
constrained framework. A lepton and a down-type quark combine into a
color triplet current to which a colored state with electric charge
$2/3$ or $4/3$ can couple. The two charge assignments of the
leptoquark correspond to fermion numbers $F=0$ and $F=2$ of the
bilinear, where $F = 3B+L$, and $B$ and $L$ are baryon and lepton
numbers (see Fig.~\ref{fig:flow}).

Our aim here is to consider one by one leptoquarks that potentially
contribute to the $b\to s \ell^+ \ell^-$ transitions, determine
correlations between effective operators affecting the $b \to s \ell^+
\ell^-$ effective Hamiltonian, and constrain the underlying couplings
from experimental data on $B_s\to \mu^+ \mu^-$, $B \to K \ell^+
\ell^-$, and $B \to X_s \mu^+ \mu^-$ decays.

\section{Effective Hamiltonian}
The effective Hamiltonian of dimension-6 at the mass scale of $b$
quark reads~\cite{Grinstein:1988me,*Misiak:1992bc,*Buras:1994dj}
\begin{align}
  \label{eq:Heff}
    \mc H_\mrm{eff} = -&\frac{4 G_F}{\sqrt{2}} \lambda_t
\Big[\sum_{i=1}^6 C_i(\mu) \mc{O}_i(\mu) \\
&+ \sum_{i=7,8,9,10,P,S} \left(C_i(\mu) \mc{O}_i(\mu) +C^\prime_i(\mu)
  \mc{O}^\prime_i(\mu) \right)\nonumber\\
&+ C_T \mc{O}_T + C_{T5} \mc{O}_{T5}\Big]\,,\nonumber
\end{align}
where $\lambda_t = V_{tb} V_{ts}^*$. Effective operators that receive
contributions from leptoquarks are the two-quark, two-lepton
operators,
\begin{align}
  \label{eq:ops}
\mc{O}_9 &= \frac{e^2}{g^2} (\bar s \gamma_\mu P_L b) (\bar \ell
\gamma^\mu \ell) \,,\\
\mc{O}_{10} &= \frac{e^2}{g^2} (\bar s \gamma_\mu P_L b) (\bar \ell
\gamma^\mu \gamma_5 \ell) \,,\nonumber\\
\mc{O}_{S} &= \frac{e^2}{16\pi^2} (\bar s P_R b) (\bar \ell \ell)\,,\nonumber\\
\mc{O}_{P} &= \frac{e^2}{16\pi^2} (\bar s P_R b) (\bar \ell
\gamma_5 \ell)\,.\nonumber\\
\mc{O}_{T} &= \frac{e^2}{16\pi^2} (\bar s \sigma^{\mu\nu} b) (\bar \ell \sigma_{\mu\nu}\ell)\,,\nonumber\\
\mc{O}_{T5} &= \frac{e^2}{16\pi^2} (\bar s \sigma^{\mu\nu} b)
(\bar \ell \sigma_{\mu\nu} \gamma_5 \ell)\,.\nonumber
\end{align}
The chirally flipped operators $\mc{O}^\prime_{9,10,S,P}$
are obtained from the above ones by $L \leftrightarrow R$ exchange.  $e =
\sqrt{4\pi \alpha}$ is the unit of electric charge, $g$ is the strong coupling,
and $P_{L,R} = (1 \mp \gamma_5)/2$.
Four-quark operators $\mc{O}_{1\ldots 6}$ and radiative penguin operators
$\mc{O}_{7,8}$ can be found in ref.~\cite{Bobeth:1999mk}. Values of
the Wilson coefficients are calculated by means of matching the full
theory onto the effective theory at the electroweak scale and subsequently
solving the renormalization group equations to run them down to scale
$\mu_b = 4.8\e{GeV}$. Decay amplitudes are conveniently expressed in terms of
effective Wilson coefficients at the scale $\mu_b$~\cite{Buras:1993xp,*Altmannshofer:2008dz},
\begin{align}
\label{eq:Ceff}
C_7^{\rm eff}(\mu_b) & =  \frac{4\pi}{\alpha_s}\, C_7 -\frac{1}{3}\, C_3 -
\frac{4}{9}\, C_4 - \frac{20}{3}\, C_5\, -\frac{80}{9}\,C_6\,,
\nonumber\\
C_9^{\rm eff}(\mu_b) & =  \frac{4\pi}{\alpha_s}\,C_9 + Y(q^2)\,,
\nonumber\\
C_{10}^{\rm eff} (\mu_b)& =  \frac{4\pi}{\alpha_s}\,C_{10}\,,\qquad
C_{7,8,9,10}^{\prime,\rm eff}(\mu_b) = \frac{4\pi}{\alpha_s}\,C^\prime_{7,8,9,10}\,,
\end{align}
where function $Y(q^2)$ was defined in~\cite{Altmannshofer:2008dz}.
For the SM contributions we will use the NNLL values
$C_7^{\eff,\mrm{SM}}(\mu_b)= -0.304$, $ C_9^{\eff,\mrm{SM}}(\mu_b) =
4.211$, and $C_{10}^{\eff,\mrm{SM}}(\mu_b) =
-4.103$~\cite{Bobeth:1999mk,*Altmannshofer:2008dz}. Numerical values
of other parameters entering theoretical predictions can be found in~\cite{Becirevic:2012fy}.

The diagrams on Fig.~\ref{fig:flow} will contribute to the Wilson
coefficients of operators~\eqref{eq:ops}.  We will assume that a
leptoquark state lies at a scale $\sim 1\e{TeV}$, still perfectly
allowed by limits set by the direct
searches~\cite{Abramowicz:2012tg,*Abazov:2006vc,*Khachatryan:2010mq,*ATLAS:2012aq,*Aad:2011ch},
where we also perform the tree-level matching. For our purposes we can
neglect the anomalous dimensions of coefficients $C_{9}^{(\prime)}$
and $C_{10}^{(\prime)}$~\cite{Bobeth:2003at}, whereas the anomalous
dimensions of scalar and pseudoscalar Wilson coefficients run with the
same anomalous dimension as $m_b(\mu)$~\cite{Logan:2000iv}. Lepton
flavor universality of all beyond the SM contributions will be assumed
throughout this work in order to make a straightforward interpretation
of experimental constraint from $\mrm{Br}(B \to K \ell^+ \ell^-)$
where a result given in~\cite{:2012vw} is a combination of $\ell = e$
and $\ell = \mu$ modes.

In the following sections we will omit the ``eff'' label when writing
down beyond the SM contributions to the effective Wilson coefficients.

\section{Observables and their Standard Model predictions}
The $B_s \to \ell^+ \ell^-$ decay branching fraction in a general NP
model reads
\begin{widetext}
\begin{align}
  \label{eq:Bsll}
 \mrm{Br}\left( B_s \to \ell^+ \ell^- \right) = &\tau_{B_s} f_{B_s}^2
 m_{B_s}^3 \frac{G_F^2 |\lambda_t|^2 \alpha^2}{(4 \pi)^3}  \beta_\ell(m_{B_s}^2) \\
& \times\left[ \frac{m_{B_s}^2}{m_b^2}\Big| C_S -C_S^\prime \Big|^2  \left(1- { 4 m_\ell^2 \over m_{B_s}^2}  \right) \right.  
   \left. +  \ \Big| {m_{B_s}  \over m_b } \left( C_P -C_P^\prime \right) + 2 {   m_\ell \over m_{B_s}} \left( C_{10} - C_{10}^{\prime  } \right) \Big|^2 \, \right] \, ,  \nonumber
\end{align}
\end{widetext}
where $\beta_\ell(q^2) = \sqrt{1-4m_\ell^2/q^2}$. The above branching
fraction is sensitive exclusively to contributions of differences
between operators with left- and right-handed quark currents, $C_{10}
- C_{10}^{\prime}$, $C_{S} -C_{S}^{\prime}$, and $C_{P} -
C_{P}^{\prime}$. The latter two combinations are
effectively constrained due to lifted helicity suppression unless the
relative phases of Wilson coefficients allow cancellations between
$C_{S}\,(C_{P})$ and
$C_S^\prime\,(C_P^\prime)$. In the SM only $C_{10}$ is present in \eqref{eq:Bsll}
and leads to prediction~\cite{Becirevic:2012fy}
\begin{equation}
  \mrm{Br}(B_s \to \mu^+ \mu^- )_\mrm{SM} = (3.3\pm 0.3)\E{-9}\,,
\end{equation}
whereas the latest $95\,\%$ confidence level bound from the LHCb experiment~\cite{PhysRevLett.108.231801} is
\begin{equation}
  \label{eq:LHCb-Bsmumu}
  \mrm{Br}(B_s \to \mu^+ \mu^- )_\mrm{exp} <4.5\E{-9}\,.
\end{equation}

The decay branching fraction, $\mathrm{Br}(B \to K \ell^+ \ell^-)$, on
the other hand, receives contributions from $C_7 + C_7^\prime$, $C_9 +
C_9^\prime$, $C_{10} + C_{10}^\prime$, $C_S + C_S^\prime$, and $C_P +
C_P^\prime$, while we have neglected contribution of the tensor
operators that have small contributions in leptoquark models, as will
be shown below. The decay width reads~\cite{Bobeth:2007dw}
\begin{equation} \label{observables-Kll}
 \Gamma(B \to K \ell^+ \ell^-)  = 2 \left( A_{\ell} + \frac{1}{3} C_{\ell} \right) \,,
\end{equation}
where $A_\ell$ corresponds to the $\theta$-independent component of
the spectrum, whereas $C_\ell$ stems from the component proportional
to $\cos^2\theta$, where $\theta$ is the angle between $\bar B$ and
$\ell^-$ in the rest frame of the lepton pair. They are expressed as
integrals over the dilepton invariant mass between $q_\mrm{min}^2 =
4m_\ell^2$  and $q^2_\mrm{max} =
(m_B-m_K)^2$,
\begin{align}
 A_{\ell} & = \int_{q_{\rm min}^2}^{q_{\rm max}^2} a_\ell \qq d
 q^2\,,\qquad C_{\ell}  = \int_{q_{\rm min}^2}^{q_{\rm max}^2} c_\ell  \qq d q^2\,.
\end{align}
The corresponding spectra are
\begin{widetext}
\begin{align}
a_\ell \qq = &\  {\cal C}(q^2)\Big[  q^2 \left( \beta_{\ell}^2(q^2) \lvert F_S(q^2)\rvert^2 + \lvert F_P(q^2) \rvert^2 \right)  +
 \frac{\lambda \qq }{4}  \left( \lvert F_A (q^2)\rvert^2 +  \lvert F_V(q^2) \rvert^2 \right)   \nn \\ 
 & \hspace{2.25cm}+ 4 m_{\ell}^2 m_B^2 \lvert F_A(q^2) \rvert^2+2m_{\ell}  \left( m_B^2 -m_K^2 +q^2\right) \text{Re}\left( F_P(q^2) F_A^{\ast}(q^2) \right)\Big]  \,,\nn \\ 
c_\ell \qq = &\  {\cal C}(q^2)\Big[ - \frac{\lambda \qq}{4} \beta_{\ell}^2(q^2) \left( \lvert F_A(q^2) \rvert^2 +  \lvert F_V(q^2) \rvert^2 \right)  \Big]\,,\nn
\end{align}
where
\begin{align}\label{F_BKell}
F_V \qq  =&   \left( C_9 + C_9^{\prime  } \right)  f_+ \qq + \frac{2 m_b}{m_B +m_K} \left(   C_7  + C_7^{\prime } \right)  f_T \qq   \,,  \nn \\ 
F_A \qq    =&  \left( C_{10} +C_{10}^{\prime  } \right)  f_+ \qq \,,   \nn\\
 F_S \qq =&     \frac{m_B^2 -m_K^2}{2m_b}  \left( C_S + C_S^{\prime} \right) f_0 \qq\,,  \nn \\
F_P \qq  =&  \frac{m_B^2 -m_K^2}{2m_b} \left( C_P + C_P^{\prime} \right)  f_0 \qq  \nn  - m_{\ell} \left( C_{10}+C_{10}^{\prime } \right) \left[  f_+ \qq - \frac{m_B^2-m_K^2}{q^2} \left(f_0  \qq - f_+ \qq \right) \right]  \,.\nn
%F_T \qq =&   \frac{2 \sqrt{\lambda \qq} \beta_{\ell}(q^2)}{m_B + m_K}  C_T f_T\qq\,,\nn\\
% F_{T5} \qq  =&  \frac{2 \sqrt{\lambda \qq} \beta_{\ell}(q^2)}{m_B + m_K}  C_{T5}f_T\qq \,. 
\end{align}
\end{widetext}
The auxiliary functions are defined as
\begin{align}
  \mc{C}(q^2) &= \frac{G_F^2 \alpha^2 \lvert \lambda_t
    \rvert^2}{512 \pi^5 m_B^3}  \beta_{\ell}(q^2)   \sqrt{\lambda \qq
  } \,,\\
  \lambda \qq &= q^4+ m_B^4 + m_K^4 -2 \left( m_B^2 m_K^2 +m_B^2 q^2 + m_K^2 q^2 \right) \,.\nn
\end{align}
Functions $F_{X}$, where $X = V,A,S,P$, corresponding to different
Lorentz structures in the effective Hamiltonian, are products of the
short distance Wilson coefficients and appropriate hadronic form
factors of $B \to K$ transition, defined as follows:
\begin{align}
\langle K(k)| \bar{s} \gamma_{\mu} b |B(p)\rangle =&  \left[ (p +
  k)_\mu - {m_B^2 - m_K^2 \over q^2} q_\mu \right] f_{+}\qq \\
&+ {m_B^2-m_K^2 \over q^2} q_{\mu} f_{0}\qq \, , \nn \\
\langle K(k)| \bar{s}\sigma_{\mu\nu}b |B(p) \rangle &= i \left( p_\mu k_\nu  - p_\nu k_\mu \right) \frac{2 f_T\qq}{m_B + m_K} \, .
\end{align}
The form factors we use were obtained by simulations of QCD on the
lattice~\cite{Zhou:2011be,*Becirevic:2012fy,*BecPrep} and using QCD
sum rules on the light
cone~\cite{Ball:2004rg,*Khodjamirian:2006st}. Details about their
parameterization and numerical values have been discussed recently
in~\cite{Becirevic:2012fy}, where the following SM prediction has been
made,
\begin{equation}
  \label{eq:BKll-SM}
 \mrm{Br} \left(B \to K\ell^+ \ell^- \right)_\mrm{SM}  = (7.0 \pm 1.8 )\times 10^{-7}\,.
\end{equation}
 Recently, BaBar experiment reported a combined measurement of
 $B^0\,(B^+) \to K^0\,(K^+) \ell^+ \ell^-$~\cite{:2012vw}
 \begin{equation}
   {\rm Br} \left(B \to K\ell^+ \ell^- \right)_{\rm BaBar} =   (4.7 \pm 0.6\pm 0.2) \times 10^{-7}  \,,
 \end{equation}
 that is compatible with the SM prediction~\eqref{eq:BKll-SM}, while
 the LHCb experiment~\cite{Aaij:2012cq} found a significantly smaller result
 for neutral $B$ decays to a muon final state
 \begin{equation}
   {\rm Br} \left(B^0 \to K^0 \mu^+ \mu^- \right)_{\rm LHCb} =
   (3.1^{+0.7}_{-0.6}) \times 10^{-7} \,.
 \end{equation}
Assuming lepton flavor universality, na\"ive average of the two
constraints gives $\mrm{Br}(B \to K \ell^+ \ell^-) = (3.8 \pm
0.6)\E{-7}$, but since the two measurements are only marginally
compatible we consider in our analysis a range of allowed values
that covers both measurements
\begin{equation}
   \label{eq:BKll-exp}
  \mrm{Br}(B \to K \ell^+ \ell^-)_\mrm{exp} = (2.5-5.5)\E{-7}\,.
\end{equation}

The inclusive decay $B \to X_s \mu^+ \mu^-$ will also play an
important role in constraining the vector operators
$C_{9,10}^{(\prime)}$. Using the formulas presented
in~\cite{Huber:2007vv,*lunghi-hiller} we get for the SM prediction in
the lower range of $q^2$
\begin{equation}
\int_{1\ \e{GeV}^2}^{6\ \e{GeV}^2}{d {\rm Br} \left(B \to X_s \mu^+ \mu^- \right) \over dq^2}dq^2 \Big|_{\rm SM}=1.59(17)  \times 10^{-6}\,,
\end{equation}
where we have kept explicit dependence on $m_{b,\mrm{pole}}^5$, contained in the
normalization factor
\begin{equation} 
  \mc{B}_0= \tau_B \ {4\
    \alpha^2\ G_F^2 |\lambda_t|^2 \ m_{b,\mrm{pole}}^5\over 3\ (4
    \pi)^5 } = 3.41(47)\times 10^{-7}\,,
\end{equation}
instead of normalizing it to the branching fraction of semileptonic $B
\to X_c \ell \nu$ decay. Leptoquark-induced additive contributions to
the above prediction will be calculated by employing formulas presented
in~\cite{Fukae:1999ww} in the approximation $m_\ell = m_s = 0$. The
partial branching ratio at low $q^2$'s has been measured at the
$B$-factories~\cite{belle-incl,*babar-incl}, resulting in an
average~\cite{Huber:2007vv},
\begin{equation}
\label{eq:incl-exp}
\int_{1\e{GeV}^2}^{6\e{GeV}^2}{d {\rm Br} \left(B \to X_s \mu^+ \mu^- \right) \over
  dq^2}dq^2 \Big|_{\rm exp}= 1.6(5) \times 10^{-6}\,.  
\end{equation}

\section{Scalars}
Scalar leptoquarks typically originate from the scalar representations
of the unification group that are required to break either the unification or the SM
gauge group. We distinguish $Q=2/3$ and $Q=4/3$ cases below.

\subsection{$Q=2/3$ scalars}
Charge $2/3$ scalar leptoquarks can couple to leptons and quarks when
their chiralities are different, therefore only $\overline{d_L} \ell_R$ or
$\overline{d_R} \ell_L$ bilinears are allowed in the interaction.
Here and in the following $d$ denotes one of the down-type quarks. The
two scalars that can form renormalizable vertices with these bilinears
transform as doublets under $SU(2)_L$,
\begin{align}
  \Delta^{(7/6)} &\equiv (3,2)_{7/6}\,,\\
  \Delta^{(1/6)} &\equiv (3,2)_{1/6}\,.\nn
\end{align}
The SM quantum numbers have been specified as $(SU(3)_c,
SU(2)_L)_Y$ and the hypercharge is defined as $Y=Q-T_3$. Both states
conserve baryon ($B$) and lepton numbers ($L$). The state
$\Delta^{(7/6)}$ will couple to the right-handed~(RH) leptons in a
gauge invariant term
\begin{equation}
  \label{eq:S23-7}
 \mc{L}^{(7/6)} =  g_R\, \overline{Q} \Delta^{(7/6)} \ell_R + \mrm{h.c.}\,,
\end{equation}
that contains a coupling of the $T_3 = -1/2$ component of
$\Delta^{(7/6)}$ to down-quarks and RH leptons. To keep the notation clean,
we have omitted flavor indices on the Yukawa couplings $g_R$ and
fields. Color indices are always contracted between the leptoquark and
the quark field. We integrate out $\Delta^{(7/6)}$ and rotate
the Yukawa couplings to the quark mass-basis by a redefinition
$D_L^\dagger g_R \to g_R$, where $D_L$ connects the mass and gauge
bases as $d_L^\mrm{gauge} = D_L d_L^\mrm{mass}$. The effective
Hamiltonian~\eqref{eq:Heff} will receive contributions to operators
with vector and
axial-vector lepton currents
\begin{equation}
  \label{eq:Delta76}
  C_9 = C_{10} = \frac{-\pi}{2\sqrt{2} G_F \lambda_t \alpha}
  \frac{(g_R)_{s\ell} (g_R)_{b\ell}^*}{M_{\Delta^{(7/6)}}^2}\,.
\end{equation}
On the other hand, the state $\Delta^{(1/6)}$ couples
via $T_3=1/2$ isospin component to the left-handed~(LH) leptons as
\begin{equation}
  \label{eq:S23-1}
 \mc{L}^{(1/6)} =  g_L\, \overline{d_R} \tilde\Delta^{(1/6) \dagger}  L  +
  \mrm{h.c.}\,,\qquad \tilde \Delta\equiv i\tau_2 \Delta^*\,.
\end{equation}
Here $\tilde\Delta$, defined with the help of the second Pauli matrix
$\tau_2$, transforms as $(\bar 3,2)_{-1/6}$. This state leaves imprint
on operators with RH quark currents and with vector and axial-vector
lepton currents
\begin{equation}
  \label{eq:Delta16}
  -C_9^\prime = C_{10}^\prime = \frac{-\pi}{2\sqrt{2} G_F \lambda_t \alpha}
  \frac{(g_L)_{s\ell} (g_L)_{b\ell}^*}{M_{\Delta^{(7/6)}}^2}\,.
\end{equation}
We have rotated the couplings to the mass basis by redefinition $D_R^\dagger g_L
\to g_L$.

Notice that scalar and pseudoscalar operators are not induced by those
two states since each of them couples exclusively either to LH or to
RH leptons whereas operators $\mc{O}_{S,P}^{(\prime)}$ involve both
lepton and quark chiralities. However, if we expand our approach and
allow for presence of both states we see that they weakly mix since
the quantum numbers of $\Delta^{(7/6)}_{T_3=-1/2}$ and
$\Delta^{(1/6)}_{T_3=+1/2}$ are equal in the broken electroweak~(EW)
phase~\cite{Hirsch:1996qy}. The mixing term at the EW scale reads
\begin{align}
 \mc{L}_\mrm{mix}^{7/6-1/6} &= \xi (H^\dagger \Delta^{(7/6)}) (H^\dagger \tilde\Delta^{(1/6)}) + \mrm{h.c.}\,,
\end{align}
where $H$ is the Higgs doublet, and $\xi$ is a dimensionless
parameter.\footnote{We have neglected the diagonal couplings to two
  Higgses, $(H^\dagger H) (\Delta^\dagger \Delta)$, with
  $\Delta=\Delta^{(7/6)}, \Delta^{(1/6)}$, that would merely shift the
  diagonal mass parameters.} The above mixing between the two
otherwise $B$ and $L$ conserving leptoquarks violates $L$ by $-2$ and $B$
by $2/3$. Radiative generation of Majorana masses for neutrinos in a
similar setting has been considered in~\cite{PhysRevD.77.055011}. The
EW symmetry breaking generates nondiagonal terms in mass matrix for
states $(\Delta^{(7/6)}_{T_3=1/2}, \Delta^{(1/6)}_{T_3=-1/2})$
\begin{equation}
  \begin{pmatrix}
    M_{\Delta^{(7/6)}}^2 & \frac{\xi^* v^2}{2} \\
   \frac{\xi v^2}{2} & M_{\Delta^{(1/6)}}^2
  \end{pmatrix}\,,
\end{equation}
where $v=246\,$GeV is the vacuum expectation value of the Higgs
field. The heavy and light mass eigenstates, $\Delta_H$, $\Delta_L$,
are mixtures of states $\Delta^{(7/6)}$ and $\Delta^{(1/6)}$
(without $T_3$ labels from now on). To illustrate consequences in
that setting let us consider a case when $M_{\Delta^{(1/6)}}
\ll M_{\Delta^{(7/6)}}$. The mass eigenstates are
\begin{equation}
  \label{eq:mixing}
  \begin{pmatrix}
    \Delta_H\\\Delta_L
  \end{pmatrix}
=
\begin{pmatrix}
  1 & \frac{\xi v^2}{2|\Delta M^2|}\\
-\frac{\xi^* v^2}{2|\Delta M^2|} & 1
\end{pmatrix}
\begin{pmatrix}
  \Delta^{(7/6)}\\\Delta^{(1/6)}
\end{pmatrix}\,,
\end{equation}
to leading order in mixing parameter, $|\xi| v^2/|\Delta M^2|$, where
$|\Delta M^2| = |M_{\Delta^{(1/6)}}^2-M_{\Delta^{(7/6)}}^2|$.
Consequently, the lighter of the two states will decrease its
mass by $|\xi| v^2/(8\sqrt{|\Delta M^2|})$ while mass of the heavier
state will increase by the same amount. In turn we generate, in
addition to $C_9^\prime$ and $C_{10}^\prime$ in \eqref{eq:Delta16}, an
entire set of scalar, pseudoscalar, and tensor operators:
\begin{align}
 \label{eq:ST23}
  C_P &= C_S = \frac{-\pi}{4\sqrt{2} G_F \lambda_t \alpha} \frac{\xi
    v^2\,(g_{R})_{s\ell}
  (g_L)^*_{b\ell}}{M_{\Delta^{(1/6)}}^2 M_{\Delta^{(7/6)}}^2} \,,\\
  -C_P^\prime &= C_S^\prime = \frac{-\pi}{4\sqrt{2} G_F \lambda_t \alpha} \frac{\xi^*
    v^2\,(g_{L})_{s\ell}
  (g_R)^*_{b\ell}}{M_{\Delta^{(1/6)}}^2 M_{\Delta^{(7/6)}}^2} \,,\nn\\
  C_T &= (C_S + C_S^\prime)/4\,,\nn\\
  C_{T5} &= (C_S - C_S^\prime)/4\,.\nn
\end{align}
Same form of expressions for the Wilson coefficients \eqref{eq:ST23} and
mixing matrix apply in the inverse mass hierarchy case, with $\Delta^{(7/6)}$
light and $\Delta^{(1/6)}$ heavy, provided we relabel $(7/6)
\leftrightarrow (1/6)$ and $\xi \leftrightarrow \xi^*$. In this case
also $C_{9}$ and $C_{10}$ of Eq.~\eqref{eq:Delta76} are present.

\subsection{$Q=4/3$ scalars}
This case corresponds to a scalar that couples to ``clashing'' fermion
flows of quark and lepton fields. Their chiralities are equal in this
case due to the well known identity $(\psi_L)^C = (\psi^C)_R$, stating
that a charge-conjugate of left-handed field transforms as a
right-handed field under the Lorentz group. Scalar bilinears that
participate in vertices are therefore $\overline{d_L^C} \ell_L$ and
$\overline{d_R^C} \ell_R$, with $\psi^C \equiv C \bar\psi^T$ and $C$
is a unitary, antisymmetric charge-conjugation matrix in spinor space.
We find a weak triplet and singlet states that couple to those
bilinears,
\begin{align}
  \Delta^{(1/3)} &\equiv  (\bar 3,3)_{1/3}\,,\\
\Delta^{(4/3)} &\equiv (\bar 3,1)_{4/3}\,.\nn
\end{align}
The isotriplet state couples exclusively to LH, whereas the
isosinglet couples to the RH fermions. They both form vertices with two
quarks which makes them baryon and lepton number violating, $B-L$
conserving leptoquarks. The isotriplet $\Delta^{(1/3)}$ interaction
with two fermionic doublets contains the relevant term involving the
$T_3 = +1$ component
\begin{align}
 \mc{L}^{(1/3)} &=\frac{g_L}{\sqrt{2}}\,\overline{Q^C} i\tau_2
 \bm{\tau}\cdot \bm{\Delta}^{(1/3)} L + \mrm{h.c.}\\
  &= g_L\, \overline{d_L^C} \ell_L \Delta^{(1/3)}_{T_3=+1} + \cdots\,.\nonumber
\end{align}
A vector of Pauli matrices $\bm{\tau}$ has been introduced. The
presence of LH fields in the above interaction implies that only
left-handed quark currents can be generated at low scale. After performing
a weak-to-mass basis transition, $D_L^T g_L \to g_L$, and integrating
out the state, we find
\begin{equation}
  C_9 = -C_{10} = \frac{\pi}{2\sqrt{2} G_F \lambda_t \alpha}
  \frac{(g_L)_{b\ell} (g_L)^*_{s\ell}}{M_{\Delta^{1/3}}^2}\,.
\end{equation}
For the isosinglet state $\Delta^{(4/3)}$ the interaction term with the
RH fermions reads
\begin{equation}
\mc{L}^{(4/3)} =  g_R \, \overline{d_R^C} \ell_R \Delta^{(4/3)} + \mrm{h.c.}\,.
\end{equation}
On the effective Hamiltonian level operators with RH quark currents
are generated
\begin{equation}
  C_9^\prime = C_{10}^\prime = \frac{\pi}{2\sqrt{2} G_F \lambda_t
    \alpha} \frac{(g_R)_{b\ell} (g_R)^*_{s\ell}}{M_{\Delta^{(4/3)}}^2}\,,
\end{equation}
where $D_R^T g_R \to g_R$ rotation has been performed along with
transition to the mass basis of fermions.

The above two scalars have same charge and can therefore mix. We can
write down the off-diagonal Higgs-induced isotriplet-isosinglet mixing
as ~\cite{Hirsch:1996qy}
\begin{equation}
\mc{L}_\mrm{mix}^{1/3-4/3} = \frac{\xi}{\sqrt{2}} \left(\tilde H^\dagger \bm{\tau} \cdot
    \bm{\Delta}^{(1/3)} H\right) \Delta^{(4/3)*} + \mrm{h.c.}\,,
\end{equation}
and find the same expression \eqref{eq:mixing} for the resulting
eigenstates, provided we replace $\Delta^{(7/6)} \to
\Delta^{(1/3)}$ and $\Delta^{(1/6)} \to \Delta^{(4/3)}$. In the limit
$M_{\Delta^{(4/3)}} \ll M_{\Delta^{(1/3)}}$ the scalar and
tensor coefficients are
\begin{align}
  \label{eq:ST43}
  C_P &= C_S = \frac{\pi}{4\sqrt{2} G_F \lambda_t \alpha} \frac{\xi
    v^2\,(g_{R})_{b\ell}
  (g_L)^*_{s\ell}}{M_{\Delta^{(4/3)}}^2 M_{\Delta^{(1/3)}}^2} \,,\\
  -C_P^\prime &= C_S^\prime = \frac{\pi}{4\sqrt{2} G_F \lambda_t \alpha} \frac{\xi^*
    v^2\,(g_{L})_{b\ell}
  (g_R)^*_{s\ell}}{M_{\Delta^{(4/3)}}^2 M_{\Delta^{(1/3)}}^2} \,,\nn\\
  -C_T &= (C_S^\prime + C_S)/4\,,\nn\\
  C_{T5} &= (C_S^\prime - C_S)/4\,.\nn
\end{align}

In conclusion, we notice that a single scalar leptoquark contributes to one of the
following 4 operators 
\begin{align}
  \mc{O}_{9}^{(\prime)} \pm \mc{O}_{10}^{(\prime)}
\end{align}
of the $b \to s \ell^+ \ell^-$ effective Hamiltonian. This is simply
due to absence of a scalar color-triplet state with couplings to both
chiralities of fermions, which are necessary to form scalar or tensor
operators. They are all \emph{chiral leptoquarks}~\cite{Leurer:1993qx}
with regard to their couplings to down-type quarks and charged
leptons. Even in the presence of two scalar leptoquarks that are
allowed to mix and thus give rise to scalar, pseudoscalar, and tensor
operators we find that Wilson coefficients corresponding to those
contributions are additionally suppressed by $v^2/M_\Delta^2$ and are
therefore less important at low energies.

%%%%%%
%%%%%% Vector leptoquarks
%%%%%%
%%%%%%
\section{Vectors}
Vector leptoquark states, if fundamental particles, are typically the
remnants of the underlying gauge bosons of the broken unification
group~\cite{Leurer:1993qx}. They can also be composite states
~\cite{Schrempp:1984nj,*Gripaios:2009dq}.

\subsection{$Q=2/3$ vectors}
Vector currents with $3B + L = 0$ always involve fermions with equal
chiralities, leading in this case to $\overline{d_L} \gamma^\mu
\ell_L$ and $\overline{d_R} \gamma^\mu \ell_R$ as the only two allowed
bilinears to which vector particles can couple to. There are two
vector leptoquarks that contain an appropriate charge 2/3 component,
\begin{align}
  V^{(3)} & \equiv (3,3)_{2/3}\,,\\
  V^{(1)} & \equiv (3,1)_{2/3}\,.\nn
\end{align}
First, the isotriplet state is $B$ and $L$ conserving and interacts with
LH fermions as
\begin{equation}
  \mc{L}^{(3)} = g_L \,\overline{Q}\, \bm{\tau}\cdot \bm{V}^{(3)}_\mu
  \gamma^\mu L + \mrm{h.c.}\,,
\end{equation}
and will, after being integrated out, contribute to the left-handed quark currents:
\begin{equation}
  C_{9} = -C_{10}  = \frac{\pi}{\sqrt{2} G_F \lambda_t \alpha}
  \frac{(g_L)_{s\ell} (g_L)^*_{b\ell}}{M^2_{V^{(3)}}}\,.
\end{equation}
Couplings have been redefined as $D_L^\dagger g_L \to g_L$. The
isosinglet state, $V^{(1)}$, on the other hand has couplings to both LH
and RH fermions, i.e. it is a \emph{nonchiral} leptoquark,
\begin{equation}
  \mc{L}^{(1)} = \left(g_L\, \overline{Q} \gamma^\mu L  +  g_R\, \overline{d_R}
    \gamma^\mu \ell_R \right)\, V^{(1)}_\mu + \mrm{h.c.}\,.
\end{equation}
In addition, $B$ is not conserved as $V^{(1)}$ can decay to two down quarks.
Because of both chiralities involved, this state contributes
to both RH and LH quark currents, as well as to scalar and
pseudoscalar operators,
\begin{align}
  C_9 &= -C_{10} = \frac{\pi}{\sqrt{2} G_F \lambda_t \alpha}
  \frac{(g_L)_{s\ell} (g_L)^*_{b\ell}}{M_{V^{(1)}}^2}\,,\\
  C_9^\prime &= C_{10}^\prime = \frac{\pi}{\sqrt{2} G_F \lambda_t
    \alpha} \frac{(g_R)_{s\ell} (g_R)^*_{b\ell}}{M_{V^{(1)}}^2}\,,\nn\\
  -C_P &= C_{S} = \frac{\sqrt{2}\pi}{G_F \lambda_t \alpha}
  \frac{(g_L)_{s\ell} (g_R)^*_{b\ell}}{M_{V^{(1)}}^2}\,,\nn\\
  C_P^\prime &= C_{S}^\prime = \frac{\sqrt{2}\pi}{G_F \lambda_t \alpha} \frac{(g_R)_{s\ell} (g_L)^*_{b\ell}}{M_{V^{(1)}}^2}\,.\nn
\end{align}

\subsection{$Q=4/3$ vectors}
Similar as in the case of $Q=2/3$ scalars, vector leptoquarks with
charge $4/3$ form vertices with quarks and leptons of different
chiralities, i.e.  $\overline{d_R^C} \gamma^\mu \ell_L$ and
$\overline{d_R^C} \gamma^\mu \ell_R$. An isodoublet state
\begin{equation}
V^{(2)} \equiv (\bar 3,2)_{5/6}\,, 
\end{equation}
induces both LH and RH lepton couplings,
\begin{align}
\mc{L}^{(2)}&= g_R\, \overline{Q^C}\, i\tau_2 V^{(2)}_\mu \gamma^\mu \ell_R +
  g_L\, \overline{d_R^C}\, \gamma^\mu\, \tilde{V}^{(2)\dagger}_\mu L +
  \mrm{h.c.}\\
&=-V_\mu^{(2),T_3=+1/2} \left[g_R\,(\overline{d_L^C} \gamma^\mu
  \ell_R)+g_L\,(\overline{d_R^C} \gamma^\mu \ell_L)\right] + \cdots\,.\nn
\end{align}
The four possible combinations of these then enter the Wilson coefficients as
\begin{align}
  C_9 &= C_{10} = \frac{-\pi}{\sqrt{2} G_F \lambda_t \alpha}
  \frac{(g_R)_{b\ell} (g_R)^*_{s\ell}}{M_{V^{(2)}}^2}\,,\\
  -C_9^\prime &= C_{10}^\prime = \frac{\pi}{\sqrt{2} G_F \lambda_t \alpha}
  \frac{(g_L)_{b\ell} (g_L)^*_{s\ell}}{M_{V^{(2)}}^2}\,,\nn\\
  C_P &= C_{S} = \frac{\sqrt{2}\pi}{G_F \lambda_t \alpha}   \frac{(g_R)_{b\ell} (g_L)^*_{s\ell}}{M_{V^{(2)}}^2}\,,\nn\\
  -C_P^\prime &= C_{S}^\prime = \frac{\sqrt{2}\pi}{G_F \lambda_t \alpha}   \frac{(g_L)_{b\ell} (g_R)^*_{s\ell}}{M_{V^{(2)}}^2}\,.\nn
\end{align}
Processes that lead to $B$ non-conservation are induced via interaction
terms of $V^{(1)}$ with two quarks
\begin{equation}
  \mc{L}^{(1)}_{qq} = \overline{Q^C}\,i\tau_2 \tilde V_\mu^{(2)}
  \gamma^\mu u_R + \mrm{h.c.}\,.
\end{equation}

\section{Constraints on leptoquark-induced effective interactions}
In each case studied in the previous sections the obtained set of
Wilson coefficients follows relations between vector and axial leptonic
currents, namely, we can always express $C_{10}^{(\prime)}$ and
$C_{P}^{(\prime)}$ with $C_{9}^{(\prime)}$ and $C_{S}^{(\prime)}$,
respectively, as
\begin{equation}
  \label{eq:partners}
\left(
  \begin{array}{r}
    C_{10}\\
    C_{10}^\prime\\
    C_P\\
    C_P^\prime
  \end{array}
\right)
  =
  \pm
\left(
  \begin{array}{r}
    C_{9}\\
   -C_{9}^\prime\\
    C_S\\   
    -C_S^\prime
  \end{array}\right)\,.
\end{equation}
Positive sign on the right-hand side applies for contributions of the
scalars $\Delta^{(7/6)}$, $\Delta^{(1/6)}$ and the vector state
$V^{(2)}$, whereas the negative sign is valid for Wilson
coefficients generated by the the scalars $\Delta^{(4/3)}$, $\Delta^{(1/3)}$,
and vectors $V^{(2)}$ and $V^{(1)}$. The contributions of the seven
leptoquark states to the effective Hamiltonian are restated in
Table~\ref{tab:LQ} where we have already employed the
identity~\eqref{eq:partners} to express all Wilson coefficients in
terms of complex $C_{10}$, $C_{10}^\prime$, $C_S$, and $C_S^\prime$
that can be chosen independently (they can be found in shaded columns of
Tab.~\ref{tab:LQ}). Because all the Wilson coefficients are invariant
under rescaling of the underlying leptoquark couplings
\begin{align}
 (g_{L,R})_{s\ell} &\to \zeta (g_{L,R})_{s\ell} \qquad (\zeta \in\mathbb{C})\,,\\ 
 (g_{L,R})_{b\ell} &\to \frac{1}{\zeta^*} (g_{L,R})_{b\ell}\,,\nonumber
\end{align}
we can further eliminate one complex degree of freedom, say $C_{10}$,
by employing
\begin{equation}
  \label{eq:relation}
 4 C_{10} C_{10}^\prime = 
   - C_{S} C_{S}^\prime\,.
\end{equation}
Only the vector states $V^{(1)}$ and $V^{(2)}$ implement the most
general framework where the current-current $\mc{O}^{(\prime)}_{9,10}$
\emph{and} scalar/pseudoscalar $\mc{O}^{(\prime)}_{S,P}$ operators are
present. Remaining states have $C_S^{(\prime)} = C_P^{(\prime)} =0$
and therefore contribute either to $C_{10}$ or $C_{10}^\prime$ (and
their $C_{9}^{(\prime)}$ partners, see eq.~\eqref{eq:partners}) as can
be seen from \eqref{eq:relation}. In fact, a combination of
(pseudo)scalar and tensor operators could also arise due to presence
of two scalar states with same electric charge, however, we have
demonstrated in the previous section those operators are further
suppressed by factor $v^2/M_{\Delta}^2$ and are therefore omitted from
Tab.~\ref{tab:LQ} and from further study. Same table also shows that
leptoquarks that conserve baryon number and therefore cannot trigger
nucleon decay~\cite{Nath:2006ut,*Dorsner:2012nq}, are limited to
contributions to operators with vector and axial-vector leptonic
currents. These states, $\Delta^{(7/6)}$, $\Delta^{(1/6)}$, and
$V^{(3)}$, can lie at or below the $1\e{TeV}$ scale and therefore
produce visible effects in $b\to s\ell^+ \ell^-$ processes. Effects of
those states and $\Delta^{(4/3)}$ are in the focus of this section. We
do not delve into study of $B$-violating vector leptoquarks that
require more thorough analysis due to presence of many operators as
well as due to their potential effect on nucleon stability.
\begin{table}[htbp]
  \centering
  \begin{tabular}[c]{||c|c|c||r|>{\color{white}\columncolor[gray]{.4}[1pt]}c|>{\color{white}\columncolor[gray]{.4}[1pt]}c|r||r|>{\color{white}\columncolor[gray]{.4}[1pt]}c|>{\color{white}\columncolor[gray]{.4}[1pt]}c|r||}
\hline\hline
S & LQ & BNC& $\mc{O}_{9}$ & $\mc{O}_{10}$ & $\mc{O}_{S}$ & $\mc{O}_{P}$ & $\mc{O}_{9}^\prime$ & $\mc{O}_{10}^\prime$ & $\mc{O}_{S}^\prime$ & $\mc{O}_{P}^\prime$\\
\hline\hline
\multirow{4}{*}{$0$} & $\Delta^{(7/6)}$ & \checkmark &$C_{10}$ &
$\boldsymbol{C_{10}}$ & & & & & &\\
 & $\Delta^{(1/6)}$ &\checkmark &&&& &$-C_{10}^\prime$ &$\boldsymbol{C_{10}^\prime}$&&\\
 &$\Delta^{(4/3)}$ && && && $C_{10}^\prime$ &
 $\boldsymbol{C_{10}^\prime}$& &\\
 &$\Delta^{(1/3)}$ && $-C_{10} $& $\boldsymbol{C_{10}}$& && &
 & &\\
\hline
\multirow{3}{*}{$1$} & $V^{(3)}$ &\checkmark & $-C_{10} $& $\boldsymbol{C_{10}}$& && &
 & &\\
 & $V^{(1)}$ && $-C_{10} $& $\boldsymbol{C_{10}}$&
 $\boldsymbol{C_{S}}$ & $-C_S$ & $C_{10}^\prime$ & $\boldsymbol{C_{10}^\prime}$
 &$\boldsymbol{C_{S}^\prime}$ & $C_{S}^\prime$ \\
 & $V^{(2)}$ && $C_{10} $& $\boldsymbol{C_{10}}$&
 $\boldsymbol{C_{S}}$ & $C_S$ & $-C_{10}^\prime$ & $\boldsymbol{C_{10}^\prime}$
 &$\boldsymbol{C_{S}^\prime}$ & $-C_{S}^\prime$\\
\hline\hline
  \end{tabular}
  \caption{Scalar and vector leptoquark
    tree-level contributions to $(\bar s b)(\bar\ell \ell)$
    effective Hamiltonian. Third column (BNC) indicates whether baryon number
    is conserved. Wilson coefficients in the shaded columns ($C_{10}$,
    $C_{10}^\prime$, $C_S$, and $C_S^\prime$)
    are taken as independent. See text for
    clarification on number of independent parameters for the last two states.}
  \label{tab:LQ}
\end{table}

\begin{figure}[!hbtp]
  \centering
  \begin{tabular}{c}
    \includegraphics[width=0.48\textwidth]{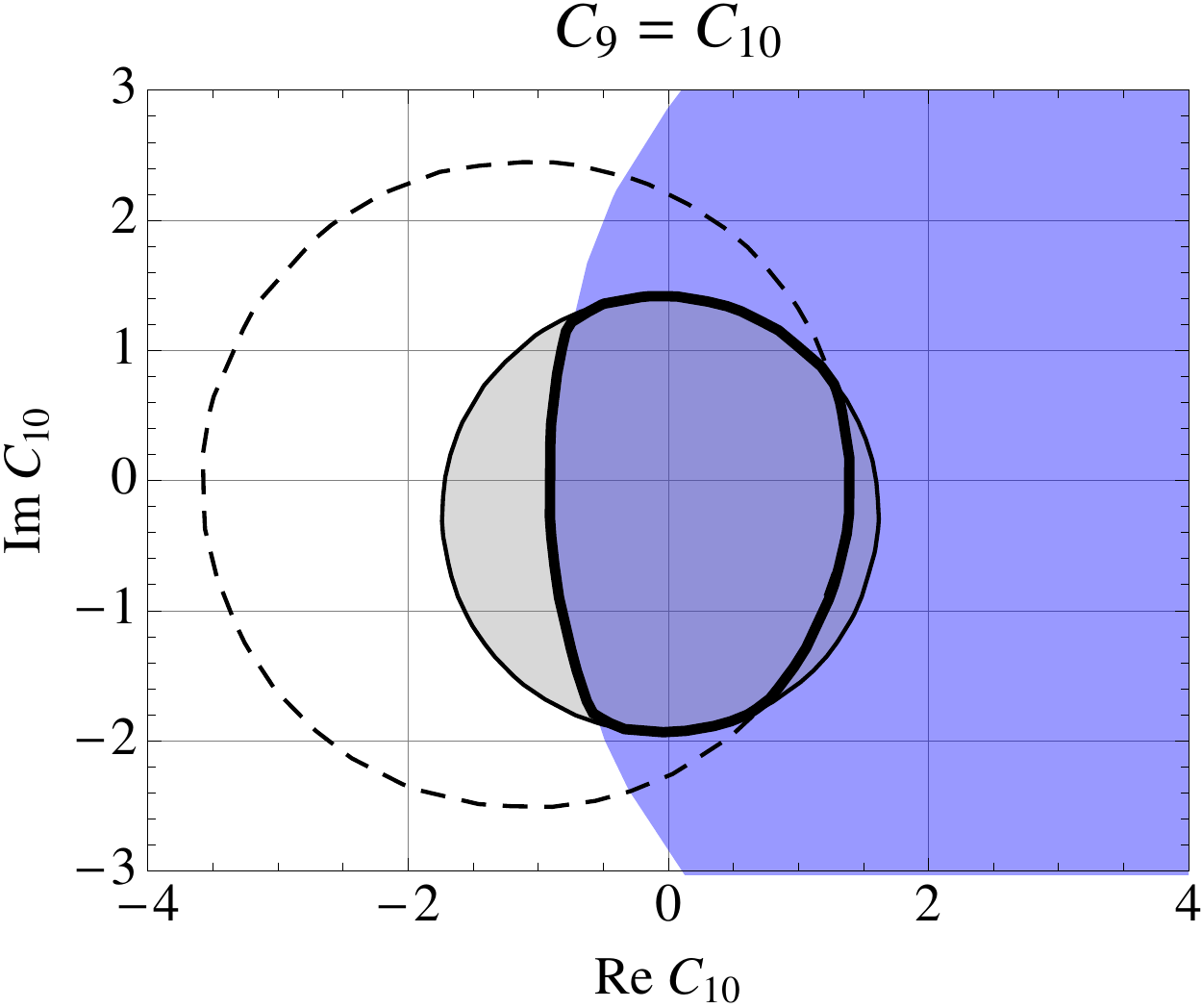} \\\\ \includegraphics[width=0.48\textwidth]{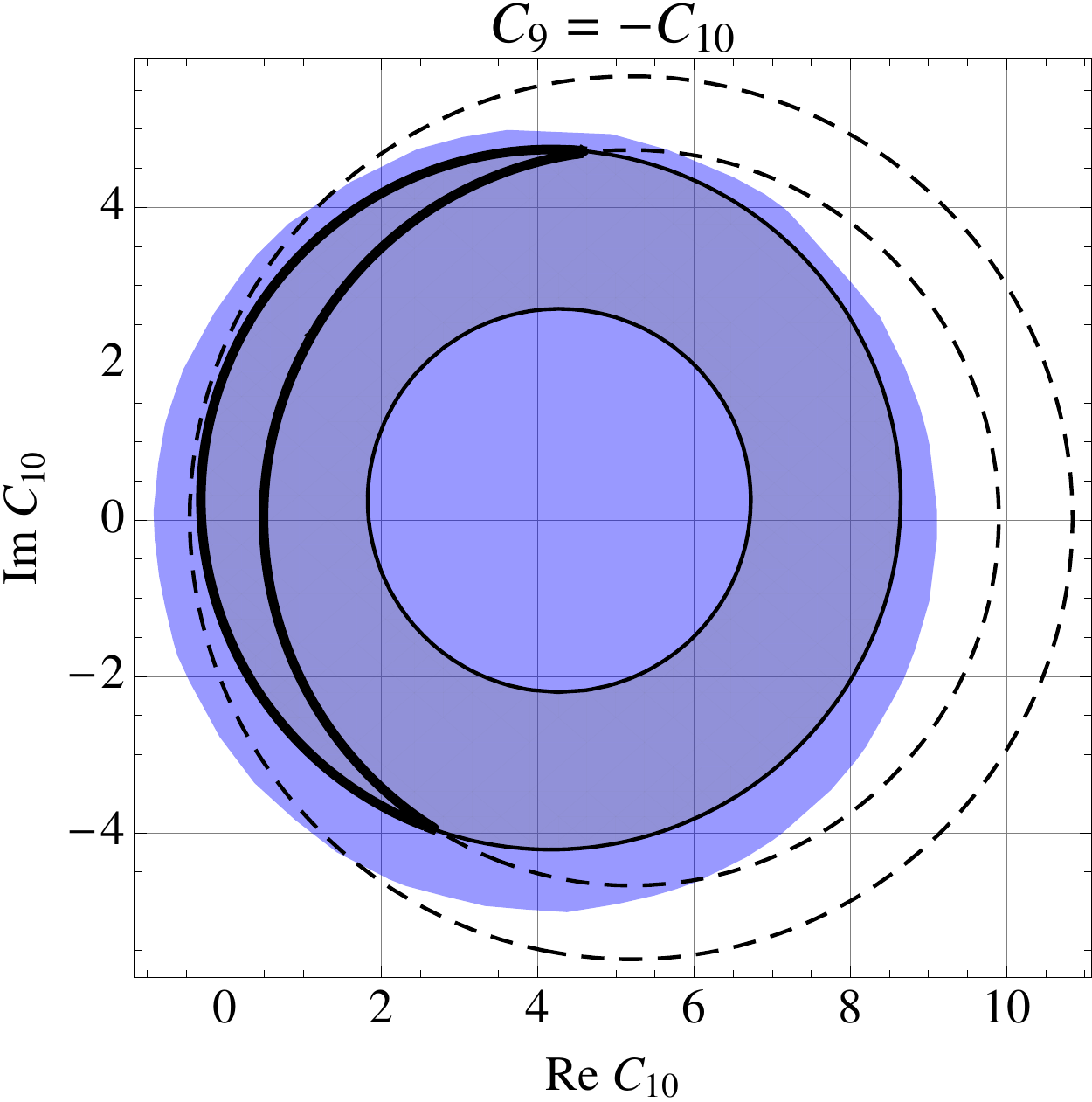}
  \end{tabular}
  \caption{Allowed regions in the complex $C_{10}$ plane in the
    leptoquark scenario where $C_9 = C_{10}$ (upper plot) or $C_9 =
    -C_{10}$ (lower plot). Blue region corresponds to $\mrm{Br}(B_s \to \mu^+
    \mu^-)$, whereas the light gray region and dashed lines mark the
    $\mrm{Br}(B \to K \mu^+\mu^-)$ and $B \to X_s \mu^+ \mu^-$
    constraints, respectively. The intersection of all three
    constraints is thickly outlined. We observe complementarity
    of the three constraints in the upper and their degeneracy in the
    lower plot.}
  \label{fig:C9C10}
\end{figure}
\subsection{$C_9 = \pm C_{10}$}
These two scenarios are realized by scalar $\Delta^{(7/6)}$ with the
$+$ sign and by vector $V^{(3)}$ with the $-$ sign. They cannot be
distinguished by the $C_9$-independent constraint $\mrm{Br}(B_s \to
\mu^+ \mu^-)$, whereas the $\mrm{Br}(B \to K \ell^+ \ell^-)$ and
partial branching fraction of $B \to X_s \mu^+ \mu^-$ decay depend
crucially on the relative sign between $C_9$ and $C_{10}$. Beyond the
SM contribution to the inclusive decay spectrum can be adapted from
formulas in ref.~\cite{Fukae:1999ww},
\begin{align}
&  \frac{d\mrm{Br}(B \to X_s \mu^+ \mu^-)}{d\hat s} = 2 \mc{B}_0
  (1-\hat s)^2\Big[ (1+2\hat s) \big\{C_{10}^\mrm{SM} \Re\left[C_{10}\right]\nn\\
   &\hspace{1cm}   \pm \Re\left[C_9^\mrm{SM}(\hat s)\, C_{10}^\ast\right] + |C_{10}|^2\big\} \mp 6C_7^\mrm{SM} \Re \left[C_{10}\right]\Big]\,,
\end{align}
where $\hat s = q^2/m_{b,\mrm{pole}}^2$ and the choice of sign should follow $C_9 = \pm C_{10}$. We show in
Fig.~\ref{fig:C9C10} how the three experimental constraints
\eqref{eq:LHCb-Bsmumu}, \eqref{eq:BKll-exp}, \eqref{eq:incl-exp}, map
onto the $C_{10}$ plane when we confront them with theoretical
predictions.
Important information in these two cases comes from the measured $B \to K
\ell^+ \ell^-$ while the effectiveness of $B \to X_s \mu^+ \mu^-$ and
the leptonic decay $B_s \to \mu^+ \mu^-$ depends on relative sign
between $C_9$ and $C_{10}$. In the $C_{9} = C_{10}$ case ($\Delta^{(7/6)}$ scalar leptoquark) the
$B\to K\ell^+ \ell^-$ decay gives the strongest constraint, however
large negative values of $C_{10}$ are effectively excluded also by
$B_s \to \mu^+ \mu^-$ due to positive interference with the SM. This
is a clear demonstration how decreasing experimental bound on
$B_s \to \mu^+ \mu^-$ is becoming more and more constraining
even for vector and axial-vector operators. The opposite relative sign
between $C_9$ and $C_{10}$ ($V^{(3)}$ vector leptoquark) allows for a
finely tuned phase of $C_{10}$ when one can effectively
cancel contributions to $\mrm{Br}(B \to K \ell^+ \ell^-)$ and $B \to
X_s \mu^+ \mu^-$. One can even decrease the two branching fractions
and therefore the lower end of the experimental predictions also
become relevant in this case.

The overlapping regions of the three constraints give for the size of
leptoquark contributions
\begin{equation}
  |C_{9,10}| \lesssim 
\left\{\begin{array}{lcl} 
4 &;& C_9 = C_{10}\\
6 &;& C_9 = -C_{10}
\end{array}\right.\,.
\end{equation}

\subsection{$C_9^\prime = \pm C_{10}^\prime$}
\begin{figure}[!hbtp]
  \centering
  \begin{tabular}{c}
    \includegraphics[width=0.48\textwidth]{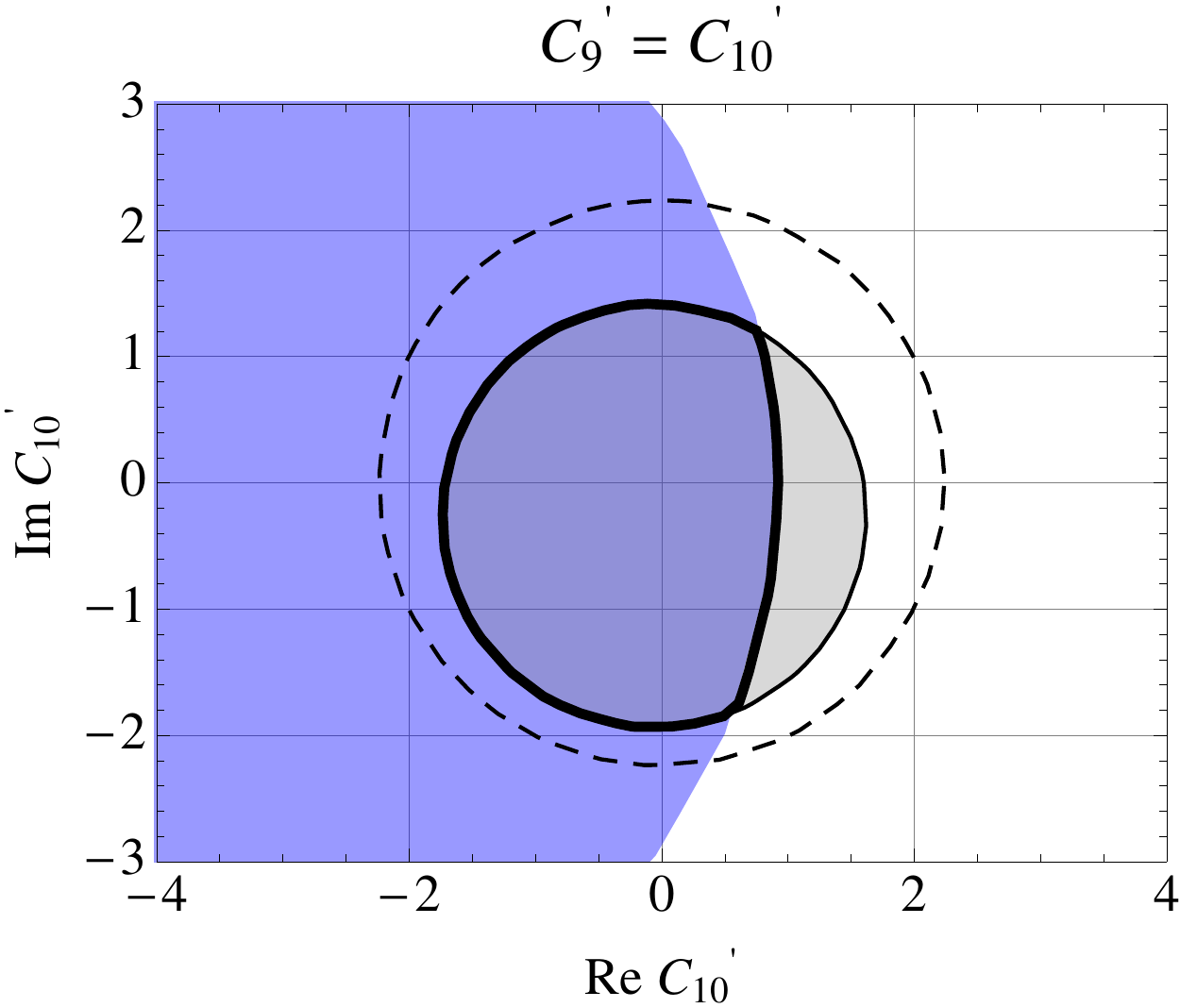} \\\\ \includegraphics[width=0.48\textwidth]{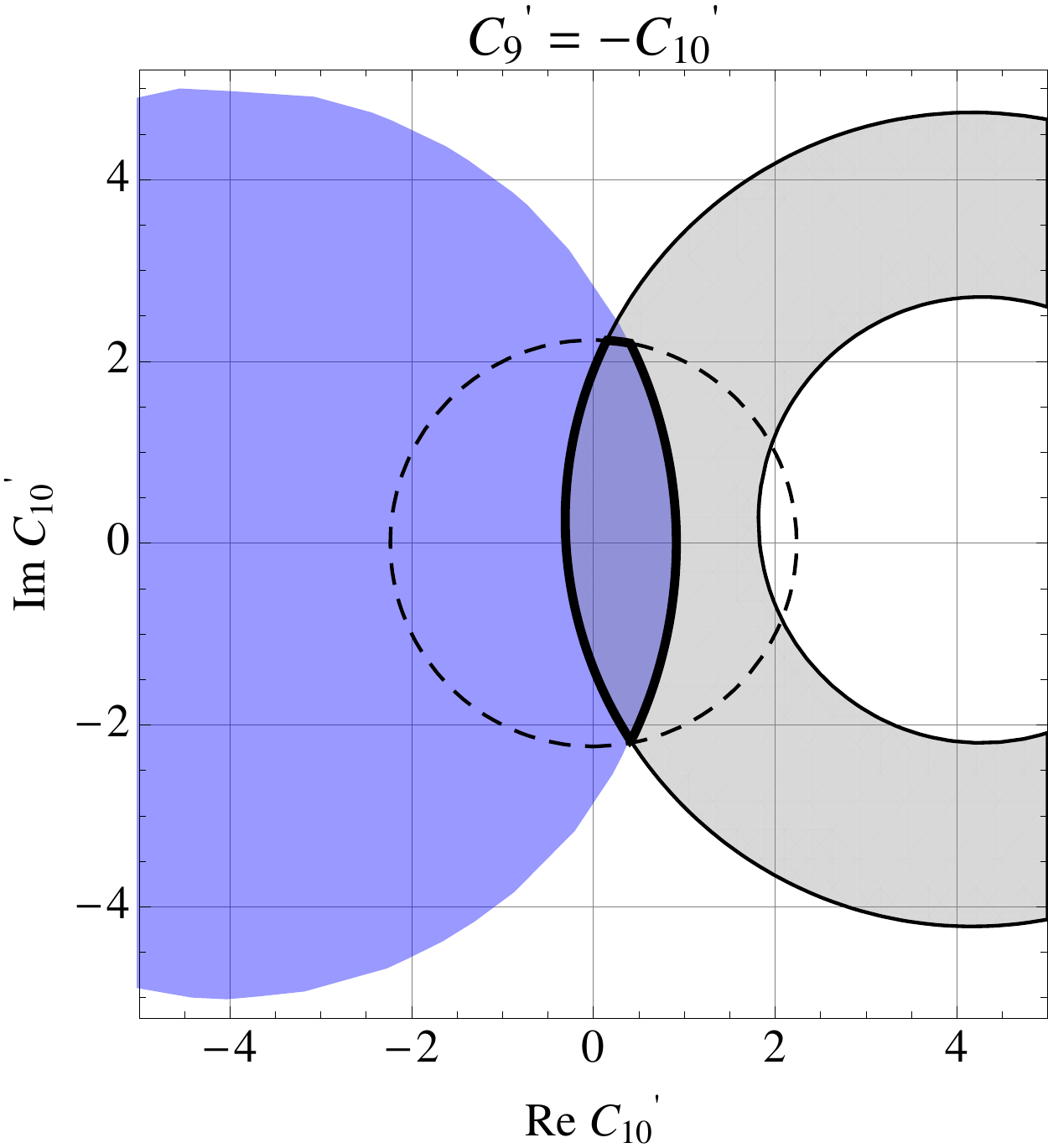}
  \end{tabular}
  \caption{Allowed regions in the complex $C_{10}^\prime$ plane in the
    scenario with $C_9^\prime = C_{10}^\prime$ (upper plot) or with $C_9^\prime =
    -C_{10}^\prime$ (lower plot). Blue region corresponds to $\mrm{Br}(B_s \to \mu^+
    \mu^-)$, whereas the gray region and dashed lines mark the
    $\mrm{Br}(B \to K \mu^+\mu^-)$ and $B \to X_s \mu^+ \mu^-$
    constraints, respectively. The intersection of all three
    constraints is outlined in thick.}
  \label{fig:C9pC10p}
\end{figure}

Scalar leptoquarks that couple to the right-handed fermions belong into
this category. States $\Delta^{(4/3)}$ and $\Delta^{(1/6)}$ will
induce such contributions with $+$ and $-$ sign,
respectively. Shift of the inclusive decay spectrum relatively to the
SM prediction can be written in these two cases as
\begin{equation}
  \frac{d\mrm{Br}(B \to X_s \mu^+ \mu^-)}{d \hat s} = 2 \mc{B}_0
  (1-\hat s)^2 (1+2\hat s) |C_{10}^\prime|^2\,.
\end{equation}
We have neglected the interference terms proportional to $m_s$ and therefore
the inclusive branching fraction is insensitive to the phase
of $C_9^\prime$. One way to distinguish the two scenarios is to
measure precisely $B \to K \ell^+ \ell^-$ that exhibits striking
sensitivity on the relative sign between $C_{9}^\prime$ and
$C_{10}^\prime$, as shown on Fig.~\ref{fig:C9pC10p}.  The allowed
regions satisfy
\begin{equation}
  |C_{9,10}^{(\prime)}| \lesssim 2\,,
\end{equation}
for both cases. However, a closer look at Fig.~\ref{fig:C9pC10p}
reveals that tension between the $\mrm{Br}(B_s \to \mu^+\mu^-)$ and
$\mrm{Br}(B \to K \ell^+ \ell^-)$ in scenario $C_9^\prime =
-C_{10}^\prime$ forces the Wilson coefficients to develop CP violating
imaginary part.  The constraint from $B_s \to \mu^+ \mu^-$ is
identical in the two cases and excludes a sizeable portion of
parameter space only in the case of flipped sign scenario
($C_{9}^\prime = - C_{10}^\prime$). On the other hand, the inclusive
decay is less sensitive to the RH current operators since the
interference terms between NP and the SM amplitude are suppressed by
$m_s$.

\section{Conclusion}
We have demonstrated in detail that color triplet bosons, i.e., leptoquarks, can generate
an entire set of effective operators of $b\to s\ell^+ \ell^-$
processes, including scalar and pseudoscalar ones. There are in total
4 scalar and 3 vector states that contribute to those operators at
tree-level.  Only two vector, baryon number violating leptoquarks are
capable of inducing (pseudo)scalar effective operators that are in
general accompanied by vector and axial-vector operators. This feature
is simply due to a fact that all scalar leptoquarks that couple to
down-type quarks and charged leptons are chiral, namely they can
couple either to right- or left-handed leptons.  This is not the case
for leptoquarks that induce $c\to u \ell^+ \ell^-$ process where a
scalar state does lead to scalar and tensor effective
operators~\cite{Fajfer:2008tm}.

Remaining 1 vector and 4 scalar leptoquarks couple to down-type quarks
and leptons chirally and their effects are limited to pairs of vector
and axial-vector effective operators. We have constrained their Wilson
coefficients by imposing the experimental constraints coming from
$\mrm{Br}(B_s \to \mu^+ \mu^-)$, $\mrm{Br}(B \to K \ell^+ \ell^-)$,
and $\mrm{Br}(B \to X_s \mu^+ \mu^-)_{[1\e{GeV}^2<q^2<6\e{GeV}^2]}$.
Importance of individual constraints depends on the particular
leptoquark state. The most constraining measurement in almost all
cases is the $\mrm{Br}(B\to K \ell^+ \ell^-)$, while $B_s \to \mu^+
\mu^-$ is also becoming a sensitive probe of (axial-)vector
operators. Presence of these operators can be tested for in transverse
asymmetries of $B \to K^* \ell^+ \ell^-$ decays as shown
in~\cite{Kruger:2005ep,*Becirevic:2011bp,*Matias:2012xw}. Finally, all
the considered leptoquark states contribute to the
electromagnetic~\cite{DescotesGenon:2011yn,*Becirevic:2012dx} and
chromomagnetic operators of both chiralities, though contributions of
this sort involve many more leptoquark couplings and are
loop-suppressed compared to the effects studied in this work.

We have found typical allowed values of leptoquark-induced Wilson
coefficients are of order $1$, which corresponds to strong
constraint $|(g_{L})_{b \ell} (g_{L})_{s\ell}|$, $|(g_{R})_{b \ell}
(g_{R})_{s\ell}| \lesssim \textrm{few}\times 10^{-2}$, if leptoquark mass is set to
$1\e{TeV}$. Note that individual $(g_{L,R})_{i\ell}$, $i=s,b$ can
still be large and allow for, e.g., explanation of the anomalous muon
magnetic moment~\cite{Dorsner:2011ai}. That very combination of
couplings also enters in direct searches for leptoquark pair
production. Consequently, final states with either two or no
$b$-quark jets are likely to be enhanced with respect to a channel
with one $b$-quark jet.

\begin{acknowledgments}
  I am indebted to D.~Be\v {c}irevi\'c who has encouraged me
  throughout the writing of this article. I thank I.~Dor\v sner and S.~Fajfer
  for reading the draft and providing constructive comments. Support by
  {\sl Agence Nationale de la Recherche}, contract LFV-CPV-LHC
  ANR-NT09-508531 is acknowledged.
\end{acknowledgments}

\bibliography{refs}

\end{document}